**Empowering the Grid: Collaborative Edge Artificial Intelligence for Decentralized Energy Systems**

Eddie de Paula Jr.[1, 2], Niel Bunda [1], Hezerul Abdul Karim [3,*], Nouar AlDahoul [4,*], Myles Joshua Toledo Tan [5,6,7,8,9,10*]

[1] Department of Computer Science, College of Computing Studies, University of St. La Salle, Bacolod, Philippines

[2] Graduate School, University of St. La Salle, Bacolod, Philippines

[3] Faculty of Engineering, Multimedia University, Cyberjaya, Selangor, Malaysia

[4] Department of Computer Science, Division of Science, New York University Abu Dhabi, Abu Dhabi, United Arab Emirates

[5] Department of Electrical and Computer Engineering, Herbert Wertheim College of Engineering, University of Florida, Gainesville, FL, United States

[6] Department of Epidemiology, College of Public Health and Health Professions and College of Medicine, University of Florida, Gainesville, FL, United States

[7] Biology Program, College of Arts and Sciences, University of St. La Salle, Bacolod, Philippines

[6] Department of Natural Sciences, College of Arts and Sciences, University of St. La Salle, Bacolod, Philippines

[8] Department of Chemical Engineering, College of Engineering and Technology, University of St. La Salle, Bacolod, Philippines

[9] Department of Electronics Engineering, College of Engineering and Technology, University of St. La Salle, Bacolod, Philippines

[10] Yo-Vivo Corporation, Bacolod, Philippines

**\* Correspondence:**
Hezerul Abdul Karim, hezerul@mmu.edu.my; Nouar AlDahoul, nouar.aldahoul@nyu.edu; Myles Joshua Toledo Tan, mylesjoshua.tan@medicine.ufl.edu

# 1. Introduction

Decentralized energy grids represent a transformative approach to energy generation and distribution. By generating power closer to demand centers and often utilizing renewable energy sources, these grids reduce transmission losses and improve energy efficiency. They leverage locally available resources such as solar and wind energy, as well as small-scale nuclear reactors, minimizing reliance on large centralized power plants and enhancing energy security (Gasca et al., 2025; Ayo-Imoru et al., 2022). This decentralized model increases resilience against outages by distributing power generation across multiple small units and balancing supply and demand more effectively through real-time electricity rates linked to grid frequency. This approach not only lowers consumer energy costs and reduces greenhouse gas emissions but also accommodates renewable energy sources and enables consumers to become "prosumers," selling surplus energy back to the grid (Bosco et al., 2022; Boyko et al., 2023).

Edge AI combines edge computing and artificial intelligence to enable data processing directly on devices at the network's edge, rather than relying on centralized cloud servers. This technology is particularly relevant in the context of decentralized energy grids. By processing data locally, Edge AI reduces transmission delays and facilitates real-time optimization of energy load and flow management, significantly enhancing the efficiency of decentralized systems (Singh & Gill, 2023; Zhang et al., 2021; Arévalo & Jurado, 2024). Autonomous control mechanisms powered by Edge AI enable decentralized decision-making for managing distributed energy resources and implementing AI-driven demand response strategies. These strategies dynamically balance supply and demand, improving grid stability (Zhang et al., 2021; Rojek et al., 2023). Furthermore, machine learning algorithms integrated with Edge AI predict equipment failures and schedule maintenance proactively, enhancing grid reliability.

This paper calls for integrating decentralized energy grids and Edge AI as essential technologies to revolutionize energy systems and address urgent global challenges. As the demand for cleaner and more resilient energy solutions continues to rise, these innovations offer practical and sustainable pathways to reshape energy generation, distribution, and consumption. By highlighting the combined strengths of decentralized grids and Edge AI, this article calls for greater commitment from policymakers, industry leaders, and stakeholders to prioritize and invest in these transformative solutions. Integrating these technologies is not just an opportunity, it is an imperative to drive meaningful progress toward a sustainable and equitable energy future.

## 2. Overview of Decentralized Energy Grids

A decentralized energy grid, or decentralized energy system, is a contemporary method of power generation and delivery that emphasizes producing electricity near the demand centers. Unlike typical centralized grids that create electricity in big plants and transmit it across extensive distances to customers, decentralized networks produce and distribute electricity closer to the consumers (Ibrar et al., 2022). It is applicable in both grid-connected and stand-alone scenarios, thereby exhibiting varying degrees of flexibility (Wynn et al., 2023). Decentralized energy grids incorporate renewable energy sources such as photovoltaic panels and wind turbines to diminish carbon emissions (Ayo-Imoru et al., 2022). Moreover, their configuration incorporates smart meters, grid frequency measurement tools, and demand response algorithms, enabling autonomous regulation of power demand and supply (Wynn et al., 2023). Demand responsive strategies enhance the equilibrium of supply and demand by incentivizing users to adjust their energy consumption away from peak hours. This strategy enhances grid efficiency and reliability. Another benefit of decentralized grids is their capacity to operate independently from the main grid, rendering them particularly advantageous for isolated or rural areas (Wynn et al., 2023).

**Table 1.** Comparison between traditional and decentralized energy systems

| Aspects | Traditional Energy Systems | Decentralized Energy Systems |
|---|---|---|
| Generation Location | Centralized, large-scale power plants | Local, near demand centers (Gasca et al., 2025; Ayo-Imoru et al., 2022; Wynn et al., 2023; Galeano Galvan et al., 2020; Shirkhani et al., 2023) |
| Energy Sources | Predominantly fossil fuels, some large-scale renewables | High integration of renewables (solar, wind, small-scale nuclear) (Ayo-Imoru et al., 2022; Wynn et al., 2023; Shirkhani et al., 2023) |
| Transmission Losses | Higher due to long-distance transmission | Lower due to proximity to demand (Wynn et al., 2023) |
| Grid Independence | Dependent on central grid | Can operate independently or in conjunction with the grid (Wynn et al., 2023; Shirkhani et al., 2023) |
| Energy Storage | Limited, large-scale storage, Primarily remote and centralized storage systems. | Distributed storage (Wynn et al., 2023; Shirkhani et al., 2023) |
| System Resilience | Vulnerable to single points of failure | More resilient due to distributed generation (Wynn et al., 2023; Shirkhani et al., 2023) |
| Technology Integration | Limited use of smart technologies | Extensive use of smart technology, AI, blockchain (Ibrar et al., 2022; Wynn et al., 2023; Galeano Galvan et al., 2020) |
| Environmental Impact | Higher carbon emissions due to fossil fuel dependency | Reduced carbon emissions through renewable energy integration (Wynn et al., 2023; Galeano Galvan et al., 2020; Shirkhani et al., 2023) |
| Economic Efficiency | Economies of scale, but higher operational and transmission costs | Potentially lower operational costs but higher initial investment in technology (Wynn et al., 2023; Galeano Galvan et al., 2020) |
| Flexibility and Scalability | Less flexible, difficult to scale quickly | Highly flexible and scalable, can be tailored to local needs (Wynn et al., 2023; Galeano Galvan et al., 2020; Shirkhani et al., 2023) |

## 3. Collaborative Edge AI in Energy Grids

### 3.1 What is Collaborative Edge AI?

Collaborative Edge AI (CEAI) is a revolutionary approach in artificial intelligence and distributed computing that incorporates edge devices to facilitate decentralized, efficient, and privacy-preserving intelligent systems. Local data processing minimizes latency, conserves bandwidth, and decreases dependence on centralized cloud infrastructure, while also mitigating privacy issues. This is accomplished using methods such as federated learning and peer-to-peer collaboration, enabling edge devices like IoT sensors, wearables, and mobile devices to jointly train AI models while preserving data locality and privacy (Abdelmoniem et al., 2024; Mughal et al., 2024).

Prominent characteristics of CEAI encompass decentralized intelligence, enabling devices to autonomously process data and render judgments, hence improving responsiveness and privacy by circumventing centralized data transfers (Badidi, 2023; Abdelmoniem et al., 2024). Inter-device communication facilitates resource and insight sharing among edge devices via peer-to-peer interactions, enhancing resource efficiency and accommodating dynamic settings such as healthcare systems and autonomous vehicles (Mughal et al., 2024; Chougule et al., 2024). Furthermore, distributed learning frameworks, including federated learning, enable devices to collaboratively train shared models, utilizing decentralized datasets to enhance global performance while maintaining privacy and scalability (Mughal et al., 2024; Abdelmoniem et al., 2024; Wynn et al., 2023).

Unlike traditional Edge AI, which operates independently and focuses on isolated tasks, CEAI emphasizes shared intelligence and cooperation among devices. Mechanisms like dynamic clustering and task allocation enable real-time responsiveness and adaptability, making it ideal for latency-sensitive applications such as healthcare diagnostics and smart transportation (Badidi, 2023; Mughal et al., 2024). By integrating distributed intelligence with robust resource and privacy management, CEAI provides a scalable and efficient framework for modern AI-driven systems.

### 3.2 How Collaborative Edge AI Applies to Energy Grids

Through predictive algorithms and real-time communication, solar panels and battery systems forecast local energy consumption, enabling efficient energy distribution based on anticipated demand and minimizing wastage (Maurya, 2024). These forecasts are shared among devices using federated learning, which trains AI models without compromising data privacy (Su et al., 2022). Devices communicate dynamically to redistribute loads, preventing localized overburdening through mechanisms like swarm-based edge computing, where nearby devices share computational and energy loads to enhance system resilience (Carnevale et al., 2022). This edge-edge collaboration balances energy usage across the network and facilitates fault detection, such as isolating power surges to prevent cascading failures, while blockchain-based decentralized energy transactions improve data transparency and anomaly detection accuracy (Methkal et al., 2024). Federated learning is instrumental in this process, enabling edge devices to share model updates instead of raw data, thereby enhancing system-wide AI models and preserving user privacy—particularly critical in urban areas with heightened privacy concerns (Abdelmoniem, 2023). Additionally, edge-oriented communication protocols like MQTT and CoAP ensure seamless and efficient real-time responsiveness, supporting energy management tasks across interconnected devices (Gong et al., 2023).

### 3.3 Benefits of Collaboration in Edge AI

Collaboration in Edge AI enhances the performance and efficiency of AI applications deployed at the network edge by offloading computation-intensive tasks to edge devices, reducing reliance on cloud resources and lowering costs associated with bandwidth and cloud computing (Oroceo et al., 2022; Zhang et al., 2019). Frameworks like the Cost Efficient Cloud Bursting Scheduler and Recommender (CECBS-R) optimize resource allocation in edge-cloud environments, promoting cost efficiency (Pasdar et al., 2021). This technology empowers small and medium-sized enterprises (SMEs) to improve operational efficiency, decision-making, and competitiveness by enabling faster reactions, personalized consumer experiences, and enhanced supply chain management (Mallela et al., 2024). In dynamic and harsh environments, such as disaster scenarios, collaborative edge computing ensures reliable and low-latency services through distributed task offloading and fault-tolerant algorithms (H.

Zhang et al., 2024). Collaboration among heterogeneous edge nodes provides ultra-reliable computing services vital for applications like autonomous vehicles and industrial IoT (M. Zhang et al., 2022). Leveraging 'small world' network dynamics enhances edge AI networks' decision-making capabilities, improving performance and scalability (Abdelmoniem et al., 2024). Federated learning architectures facilitate incremental training of machine learning models without heavy dependence on centralized resources, further advancing scalable and flexible deployment of services across geo-distributed infrastructures (Flores, 2020; M. Zhang et al., 2022). Additionally, edge AI bolsters security applications by enabling real-time monitoring and threat detection while ensuring data privacy through local processing of sensitive information (Mohamed & Al-Jaroodi, 2023).

## 4. Applications of Collaborative Edge AI

Collaborative Edge AI (CEAI) plays a transformative role in optimizing energy systems across various domains, leveraging advanced AI techniques and distributed computing resources. In renewable energy systems (RES), CEAI utilizes machine learning algorithms to predict energy production and consumption patterns, enhancing supply-demand balance and supporting grid stability through dynamic demand response strategies (Raman et al., 2024; Ukoba et al., 2024; Swarnkar et al., 2023). Smart grids powered by AI autonomously manage energy distribution, fault detection, and operations, improving efficiency and reliability (Ukoba et al., 2024; Verma et al., 2024). Federated learning frameworks enable decentralized energy systems, such as microgrids, to optimize load management and energy distribution while preserving data privacy (Hamidi et al., 2023; Nikbakht et al., 2024).

CEAI further enhances energy storage systems by predicting degradation patterns, improving energy retention, and facilitating the integration of renewable sources like solar PV and wind (Basha & Jubilson, 2024; Swarnkar et al., 2023). Predictive maintenance models powered by AI reduce downtime and maintenance costs by forecasting equipment failures, thereby improving the efficiency and reliability of RES (Swarnkar et al., 2023). Additionally, combining AI with blockchain technology enables secure and transparent energy trading through smart contracts, fostering collaborative energy management among decentralized units like smart homes and electric vehicles (Nikbakht et al., 2024; Wang & Ben Abdallah, 2022).

In microgrid operations, CEAI employs multi-agent deep reinforcement learning (MADRL) and Conditional Value-at-Risk (CVaR) to address uncertainties in energy consumption and generation, ensuring optimal energy scheduling (Munir et al., 2019, 2021). Edge-cloud collaborative architectures enhance fault diagnosis by processing data locally with advanced techniques such as deep learning on infrared thermal imaging (IRT), improving accuracy and efficiency (Chen et al., 2024). Multi-agent systems, including algorithms like multi-agent proximal policy optimization (MAPPO) and hierarchical multi-agent deep reinforcement learning (HMADRL), optimize energy interactions between microgrids and simplify control complexities, enhancing the scalability and resilience of energy systems (X. Xu et al., 2024).

In smart buildings, CEAI improves energy usage by managing systems like HVAC and lighting, leveraging real-time data processing to reduce waste and operational costs (Himeur et al., 2022; Yuan, 2022). Similarly, industrial applications utilize CEAI for automated manufacturing processes, optimizing operations based on predictive analytics to achieve significant energy savings (Goel, 2014; Soret et al., 2022). CEAI also supports transportation through route optimization, energy-efficient driving, and predictive maintenance for electric vehicles and drones (SaberiKamarposhti et al., 2024; Soret et al., 2022). In agriculture, it facilitates resource-efficient practices such as precision irrigation, reducing energy consumption and promoting sustainable farming (C. Huang et al., 2022).

Federated learning and hybrid edge-cloud architectures further enhance CEAI's effectiveness, allowing localized data processing for real-time anomaly detection, energy management, and reduced latency while maintaining data privacy (Su et al., 2022; Zhai et al., 2021; Himeur et al., 2022). AI-driven predictive maintenance and anomaly detection models improve fault resilience across energy systems, including smart grids and HVAC, by enabling early fault detection and self-healing mechanisms (Martinez-Viol et al., 2020; Yussuf & Asfour, 2024). These advancements ensure reliability, efficiency, and sustainability in energy systems.

## 5. Technical Challenges and Solutions

### 5.1 Data Privacy and Security

Resolving data privacy and security issues in decentralized energy networks necessitates a comprehensive strategy that integrates sophisticated technology solutions. Encryption methodologies such as homomorphic encryption and secure multi-party computation allow data to remain encrypted while processing, thereby safeguarding sensitive information in collaborative settings. This is particularly significant in Edge AI systems, as data is disseminated across numerous devices and susceptible to interception during transmission. These encryption techniques protect user data while allowing essential processes, such as AI model training, to proceed securely (Zhang et al., 2021). Likewise, the implementation of differential privacy, which adds noise to data or model updates, further conceals individual data contributions, hence diminishing the likelihood of sensitive information being deduced during collaborative learning processes (Mughal et al., 2024; Ouyang et al., 2024).

Decentralized authentication solutions, including those utilizing blockchain and distributed ledger technology, improve the integrity and transparency of device interactions and data transfers inside decentralized grids. Blockchain guarantees that all transactions and communications are securely recorded and resilient to tampering, hence enhancing trust among devices and mitigating vulnerabilities to fraud or cyberattacks (Wynn et al., 2023; Badidi, 2023). Furthermore, federated learning frameworks are essential for preserving data privacy by allowing edge devices to collectively train AI models without disclosing raw data. This method reduces data exposure while preserving the efficacy of model training. To augment security, approaches such as secure aggregation and cryptographic procedures can be employed to guarantee that the shared model updates stay confidential. Collectively, these technologies establish a formidable safeguard against privacy violations and security risks, allowing Edge AI systems to enhance grid efficiency while preserving consumer confidence and adhering to regulatory standards.

### 5.2 Scalability and Integration

The implementation of Edge AI in decentralized energy grids encounters critical challenges related to scalability, integration, device heterogeneity, and resource constraints. Scalability issues arise from the growing number of edge devices, such as IoT sensors and smart meters, that generate massive volumes of real-time data requiring local processing. The dynamic nature of energy grids further complicates this, as renewable energy sources like solar and wind exhibit variability, demanding real-time analytics and decision-making. Device heterogeneity exacerbates the challenge, as devices with different communication protocols, computational capacities, and architectures must seamlessly operate together. Additionally, resource constraints such as limited bandwidth, energy, and processing power at the edge nodes hinder the efficient deployment of computationally intensive AI models. These factors collectively create bottlenecks in real-time data processing and integration, complicating system-wide interoperability and performance optimization (Mughal et al., 2024; Badidi, 2023; Wynn et al., 2023)

To overcome these challenges, adopting a hierarchical and modular Edge AI architecture can enhance scalability and address device heterogeneity. For instance, multi-edge clustering frameworks like MEC-AI HetFL can group devices with similar characteristics, enabling localized processing and reducing the computational burden on individual nodes (Mughal et al., 2024). Federated learning (FL) further mitigates resource constraints by facilitating decentralized model training while preserving data privacy, minimizing the need for data transfer and central computation (Abdelmoniem et al., 2024). Integration challenges can be addressed by implementing open standards and protocols, such as OpenADR and interoperable middleware platforms, to unify communication across heterogeneous devices (Wynn et al., 2023). Dynamic resource allocation strategies and techniques like edge caching and adaptive scheduling can optimize resource utilization, ensuring efficient data processing under constrained environments. These solutions collectively enhance the scalability and integration of Edge AI systems in decentralized energy grids while mitigating the effects of device heterogeneity and resource limitations (Mughal et al., 2024; Badidi, 2023)

## 6. Current Deployments

A key component of decentralized energy systems, virtual power plants (VPPs) integrate distributed energy resources (DERs), such as solar panels, wind turbines, battery storage, and demand-response mechanisms, into a single, coordinated network (Islam et al., 2024; Ullah et al., 2024). Unlike traditional centralized power plants, VPPs offer enhanced flexibility, real-time grid balancing, and active participation in dynamic energy markets (Xie et al., 2024). According to Islam et al. (2024), AI-driven solutions are increasingly used to address operational challenges in decentralized energy systems, including energy intermittency, market price volatility, load uncertainty, and cybersecurity threats (Venegas-Zarama et al., 2022).

One notable example is Tesla's VPP in South Australia, launched in 2018. It connects thousands of Tesla Powerwall batteries in residential homes to stabilize the grid, optimize power flows, and prevent outages (Breck & Link, 2018). AI-driven predictive analytics are employed to manage the variability of renewable sources by forecasting supply-demand imbalances and dynamically adjusting energy distribution

(Shabanzadeh et al., 2016). This model demonstrates AI's potential in improving both grid stability and energy efficiency (Cao et al., 2021).

Another case is TEPCO's decentralized energy system in Tokyo, developed after the Fukushima nuclear disaster to strengthen disaster resilience. Operational since 2017, the system integrates rooftop solar panels, battery storage, and EV charging stations to enhance energy security and distribution optimization (Ullah et al., 2024). Edge AI plays a crucial role by enabling localized energy storage and real-time grid balancing, allowing for rapid power restoration during emergencies. AI-powered peak shaving helps reduce demand stress during Tokyo's peak usage periods, maintaining grid stability (Shabanzadeh et al., 2015).

Centrica's VPP in Europe represents one of the most advanced uses of Edge AI in decentralized energy markets (Venegas-Zarama et al., 2022). Spanning several countries, this VPP aggregates renewable assets, including solar installations, wind farms, and battery systems, to actively trade energy in real time (Zurborg, 2010). Edge AI-powered market prediction algorithms optimize energy transactions by responding to real-time price changes (Cao et al., 2021). Additionally, AI-based automated dispatch of distributed energy resources has cut operational costs. However, the integration of heterogeneous energy sources across regions poses scalability challenges (Qin et al., 2021). Moreover, increased automation brings heightened cybersecurity risks, necessitating robust encryption and authentication protocols (Antonopoulos et al., 2021).

## 7. Future Directions

The future of CEAI is poised for transformative growth, driven by technologies like blockchain and machine learning (ML) models. Blockchain's decentralized and immutable ledger improves data security and privacy in ML applications, enabling secure transactions without intermediaries (Akrami et al., 2023; Ural & Yoshigoe, 2023). Techniques like federated learning FL and proof of learning leverage blockchain to support privacy-preserving data sharing and reliable model validation, fostering decentralized AI systems (Ahmed & Alabi, 2024; Miglani & Kumar, 2021; Ural & Yoshigoe, 2023). Privacy-aware model training across distributed devices addresses communication and security challenges (Yu et al., 2023). The use of 5G/6G networks supports low-latency, high-speed data transfer for real-time applications, such as autonomous systems, while reinforcement learning enhances adaptive decision-making in dynamic environments (Himeur et al., 2024). Blockchain is also transforming supply chain management by improving transparency, security, and efficiency, despite challenges with scalability and energy consumption (Himeur et al., 2024; Miglani & Kumar, 2021; Ural & Yoshigoe, 2023; Yu et al., 2023).

Energy inefficiency remains a significant obstacle to sustainable CEAI deployment. Innovative solutions like immersion cooling and federated management are essential for improving energy efficiency and enabling sustainable growth (Arroba et al., 2024). Public engagement in AI policymaking is limited, particularly regarding ethical issues. Policymakers must encourage public participation and adopt ethical frameworks to align AI technologies with societal needs (Schiff, 2024).

Governance frameworks for CEAI must blend innovation with oversight, treating AI as an active participant in regulatory structures. Inclusive governance involving governments, civil society, and the private sector is crucial for balanced, transparent decision-making, particularly in healthcare, where complex ethical and legal challenges arise (Mazzi et al., 2023; Karim & Vyas, 2023; Ulnicane et al., 2021). Combining federated learning with blockchain strengthens data security and privacy by decentralizing processing and ensuring integrity (Yang et al., 2023). Governance models need to be flexible and adaptive, evolving alongside technological advancements. Starting with focused models and scaling them ensures effective management of new challenges (Sepasspour, 2023). International collaboration is vital, with science diplomacy and global frameworks facilitating harmonized regulations (Ulnicane et al., 2021). Stakeholder collaboration platforms enhance governance in public services and policymaking (Chen et al., 2024).

Innovations like the Collaborative Deep Neural Network (DNN) model selection scheme improve edge server efficiency, while information fusion drives advancements in edge intelligence (A. Xu et al., 2024; Zhang et al., 2022). Interdisciplinary research also promotes ethics-by-design and value-sensitive design frameworks, addressing societal impacts (Bisconti et al., 2023). Bridging cognitive gaps, interdisciplinary approaches foster trust, transparency, and ethics in human-machine collaboration, critical for fields like healthcare and sustainability (R. Xu et al., 2019; Oberer & Erkollar, 2023).

## 8. Conclusion

In conclusion, the integration of decentralized energy grids with collaborative Edge AI represents a transformative shift in energy management. By processing data locally and leveraging AI-driven decision-making, Edge AI enhances grid efficiency, resilience, and sustainability (Singh & Gill, 2023; Zhang et al., 2021). This approach enables real-time energy load optimization, predictive maintenance, and decentralized control, reducing transmission losses and dependency on centralized infrastructure (Maurya, 2024; Wynn et al., 2023). Furthermore, federated learning ensures data privacy while allowing edge devices to improve energy forecasting and management collaboratively (Su et al., 2022; Abdelmoniem et al., 2024). The synergy between these technologies not only optimizes renewable energy use but also empowers consumers to participate actively in the energy market, fostering a more distributed and intelligent power system (Boyko et al., 2023; Bosco et al., 2022).

To realize the full potential of collaborative Edge AI in energy grids, policymakers, industry leaders, and researchers must take proactive steps toward widespread adoption. Investment in Edge AI infrastructure, regulatory frameworks supporting decentralized energy systems, and interdisciplinary collaborations will be essential in overcoming technical and implementation challenges (Badidi, 2023; Mughal et al., 2024). Stakeholders must advocate for standardization and interoperability to ensure seamless integration of AI-driven solutions across different energy networks (Wynn et al., 2023; Zhang et al., 2024). Additionally, public and private sector partnerships should drive innovation and incentivize research into privacy-preserving AI models, adaptive control mechanisms, and energy-efficient AI deployment strategies (Methkal et al., 2024; Oroceo et al., 2022).

Looking ahead, the future of energy systems will be characterized by autonomous, self-optimizing networks that respond dynamically to real-time conditions. With advances in AI, blockchain, and 5G/6G networks, energy grids will become more adaptive, secure, and democratized (Himeur et al., 2024; Yang et al., 2023). The vision is a resilient, low-carbon, and intelligent energy ecosystem where decentralized power generation, AI-driven automation, and collaborative intelligence work in unison to achieve energy security and sustainability (Nikbakht et al., 2024; Raman et al., 2024). Embracing collaborative Edge AI is not just an option. It is an imperative step toward a smarter and greener future.